\documentstyle[aps,prl]{revtex}
\begin{document}
\title{\vspace*{-5mm}\hfill
 {\normalsize }
% {\normalsize submitted to Phys. Rev. Lett.}
        \vspace{5mm}\\
Structure and stability of a high-coverage (1$\times$1) oxygen phase
 on Ru\,(0001)}
\author{C. Stampfl, S. Schwegmann, H. Over, M. Scheffler, and G. Ertl}
\address{
Fritz-Haber-Institut der Max-Planck-Gesellschaft,
Faradayweg 4-6, D-14\,195 Berlin-Dahlem, Germany}
\date{Received 27 February 1996}
%\twocolumn[
\maketitle
\begin{quote}
\parbox{16cm}{\small
The formation of chemisorbed O-phases on Ru\,(0001) by exposure to O$_{2}$ at 
low pressures is apparently limited to coverages $\Theta \leq 0.5$. 
Using low-energy electron diffraction and density functional theory
we show that this restriction is caused by kinetic hindering
and that a dense  O overlayer ($\Theta = 1$) can be formed with
a (1$\times$1) periodicity.
The structural and energetic properties of this new adsorbate phase
are analyzed and discussed in view of attempts to bridge the so-called
``pressure gap'' in heterogeneous catalysis. It is argued that
the identified system actuates
the unusually high rate of oxidizing reactions at Ru surfaces under high
oxygen pressure conditions.
}
\end{quote}
%]
PACS numbers: 68.35.Bs, 61.14.Hg

%\newpage
\vspace{0.5cm}
Chemisorption studies with well-defined single crystal surfaces are 
propelled by the prospect to gain deeper insight into the 
elementary steps 
governing heterogeneously catalyzed reactions. Experiments of this kind are 
usually conducted under ultrahigh vacuum (UHV) conditions with partial 
pressures typically $\leq$ 10$^{-6}$ mbar 
in order to apply the standard tools of 
surface physics and to control the state of the surface. In contrast, ``real'' 
catalytic reactions take place at much higher pressures (frequently even 
exceeding 1 bar); this difference is sometimes referred to 
as the ``pressure gap''.

Although there exists a number of examples for which extrapolation of data 
over a wide pressure range has been demonstrated to be safely justified 
\cite{topsoe}, 
such a conclusion may by no means be generalized. The concentrations of 
certain surface species relevant for catalysis at high pressures may be so 
small under low pressure conditions that they escape detection, either for 
thermodynamic (low stability and hence short surface residence time) or for 
kinetic reasons (low sticking coefficient). Such an example is presumably 
the Ru\,(0001) surface interacting with O$_{2}$: Dissociative 
chemisorption under low-pressure conditions leads to apparent saturation 
at a coverage $\Theta = 0.5$ (the coverage $\Theta$ 
being defined as the ratio of the
concentration of adparticles to that of substrate atoms in the topmost 
layer) associated with the formation of an ordered (2$\times$1)-superstructure \cite{madey,surnev}. 
% In the unit cell of this phase the mutual separation between neighboring O 
% atoms is two and one Ru lattice constants, respectively, so that -- at least
% from purely geometrical arguments -- also a 
% (1$\times$1)-phase with $\Theta = 1$ 
% in which the O atoms form the same lattice as the Ru\,(0001)-substrate 
% seems to be feasible.
Since the smallest separation of O atoms in the (2$\times$1)-O overlayer is
only one lattice constant, also a (1$\times$1)-O phase with $\Theta =1$ in
which the O atoms form the same lattice as the Ru\,(0001) seems to be feasible
(at least from purely geometrical arguments). Indeed, from high pressure experiments of 
the catalytic oxidation of carbon monoxide on Ru\,(0001) the formation of a 
high-coverage O-phase was 
speculated \cite{peden}, and recent
density-functional theory (DFT) calculations suggested the existence of
an ordered O adlayer with  $\Theta = 1$, for which the
oxygen is still well chemisorbed, but with a much weaker bond
than for the other, more open phases, making the $\Theta = 1$
adlayer particularly effective for oxidizing reactions~\cite{stampfl}.
An experimental proof of this theoretical result, and in
particular, an analysis of whether or not oxygen might be present
below the surface Ru layer is still missing.
If the formation of the high coverage phase were indeed only kinetically
obstructed, then it may be realized by high oxygen pressure or by
exposure to a more strongly oxidizing
molecule such as 
NO$_{2}$. The latter approach was first employed to reach high O coverages on 
Pt\,(111) \cite{segner,banse}, and more recently it was demonstrated that also
for Ru\,(0001) a concentration of surface oxygen can be obtained
reaching, or even exceeding,
$\Theta = 1$~\cite{weinberg,malik}. 

In the present letter we report the confirmation of the formation of a 
high-coverage $\Theta = 1$ oxygen phase on Ru\,(0001) where, through the 
methods 
of low-energy electron diffraction (LEED) and density-functional theory (DFT)
calculations, we determined the 
detailed atomic geometry and the associated energetics which give insight 
into the behavior of O on Ru\,(0001) in general.
Using the technique of dissociative NO$_{2}$ chemisorption, 
we show that this phase is indeed an on-surface $(1 \times 1)$ adlayer
with a negligible amount of subsurface oxygen.

The experiments were conducted in a UHV chamber 
(base pressure $= 1 \times 10^{-10}$ mbar) 
equipped with a display type four grid LEED optics and 
with standard facilities for surface cleaning and characterization. Details 
about the experimental set-up and sample preparation can be found 
elsewhere \cite{over}.
 The preparation of the oxygen overlayer with $\Theta = 1$ was 
accomplished by first exposing the Ru\,(0001) surface to O$_{2}$ at room 
temperature until saturation was reached, i.e. the (2$\times$1)-O 
phase was formed. Subsequently, the Ru\,(0001)-(2$\times$1)-O
surface was exposed for 15 minutes to
NO$_{2}$ at a pressure of $5 \times 10^{-7} $ mbar.
The sample temperature was chosen now to be 
600\,K so that during decomposition of NO$_{2}$ into adsorbed O and NO (the 
dissociative sticking coefficient of NO$_{2}$ is about one), only O remained
on the surface while NO was instantly released into the gas phase, thus 
enriching the oxygen content at the surface; note that NO desorption from 
Ru\,(0001) is completed at about 500\,K \cite{hayden}.
Concomitant
LEED measurements 
indicated at first a sharpening of the (2$\times$2)
pattern followed by a gradual 
transformation into a bright (1$\times$1)-pattern with low back-ground
intensity, 
consistent with the formation of a (1$\times$1)-O phase. This 
(1$\times$1)-O structure 
remained stable even upon further NO$_{2}$ exposure so that no additional 
oxygen accumulated at the surface, either because the sticking 
coefficient dropped or the excess oxygen atoms disappear into the 
subsurface region. Subsequently recorded thermal desorption spectra of O, 
in combination with the structural analysis (described below), give 
compelling evidence that additional O atoms are indeed formed which 
dissolve into the bulk of the sample, leaving the oxygen concentration in the 
(1$\times$1)-O overlayer constant. A similar behavior was also observed by 
Weinberg \cite{weinberg}. LEED intensity measurements were performed at normal 
incidence of the primary beam at a sample temperature of 100\,K. A 
computer-controlled video camera was used to record integrated spot 
intensities of five integral-order beams (energy range 50 to 620\,eV) from the 
fluorescence screen. LEED curves were calculated using the program 
code of Moritz \cite{moritz}
and compared with the experimental LEED curves 
by applying an automated least-squares optimization scheme \cite{kleinle},
based either on the reliability ($R$) factors $R_{\rm DE}$~\cite{kleinle2} or
$R_{\rm P}$~\cite{pendry}.

For the DFT calculations we employed the 
generalized gradient approximation (GGA) of Perdew {\em et\, al.}~\cite{perdew}
for the exchange-correlation functional and used {\em ab initio}, 
fully separable pseudopotentials.
The surface was 
modeled by a supercell consisting of four layers of Ru\,(0001)
where the O atoms were adsorbed on one side of this slab
\cite{neugebauer,stumpf}.
We relaxed the positions of the O atoms and of the Ru atoms in the top two
layers, keeping the lower two Ru layer spacings 
fixed at the bulk values. Details of the calculations, as well as a discussion
of the clean Ru surface and more open O adlayers, can be found in
Ref.~\cite{stampfl}.

In the LEED analysis, the O-Ru layer spacing and the first two Ru-Ru 
layer spacings of the Ru\,(0001)-(1$\times$1)-O surface were simultaneously
refined, 
starting the automated search from different adsorption sites of oxygen, 
namely the fcc-, hcp-, on-top- and bridge-sites. The resulting optimum 
$R$-factors for the quoted adsorption sites are compiled in Tab.~I from
which it 
becomes evident that the hcp-position is clearly favored. In addition to these 
standard configurations, we tested a (1$\times$1)
 structure placing O in both the fcc 
and hcp position with variable concentrations. The mixing of adsorption 
sites with variable concentrations was simulated within the framework of 
the averaged-t-matrix approximation \cite{jona}.
 The multiple scattering between 
neighboring O-atoms in fcc and hcp-sites could alternatively be switched 
off. The best agreement with the experiment was in all cases achieved by the 
adsorption of O atoms exclusively in hcp-sites. Even a small concentration 
($\Theta = 0.1$) of O-atoms in fcc sites deteriorated the theory-experiment 
agreement appreciably, as indicated by an increase of the $R$-factors by 0.05. 
The next issue we addressed carefully was as to whether the results are 
sensitive to the presence of sub-surface oxygen. For this purpose we put, in 
addition to the O-(1$\times$1)
 overlayer, O atoms, with variable concentrations, 
into the octahedral sites between the first two Ru-layers and optimized their 
concentration and the structural parameters. The optimization scheme 
always ended with a structure in which O adsorbs again solely in hcp-sites 
on the surface. 
On the 
basis of these simulations we can therefore safely rule out any O
concentration exceeding 0.1 monolayer (ML)
between the top two Ru layers. Thermal desorption 
spectra (TDS) of oxygen, on the other hand, indicate that there is much more 
oxygen than $\Theta = 1$ present in this system and that appreciable amounts
of 
oxygen are still present on the surface even after many thermal desorption 
cycles. Obviously, these O atoms prefer to dissolve into the bulk rather than 
to form a subsurface phase between the first two Ru-layers. 

\begin{figure}
\begin{picture}(-300,200)(100,250) 
\put(0,-150) { \includegraphics{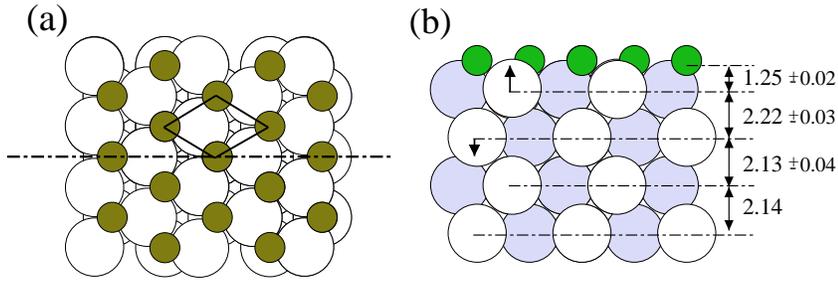}} 
\end{picture}
\caption{Top view (a) and side view (b) of the atomic geometry of 
(1$\times$1)-O/Ru\,(0001) with O sitting in the hcp-hollow site. The 
arrows indicate the direction of the displacements of the substrate 
atoms with respect to the bulk terminated positions. The distances 
are those determined by LEED and are given in \AA.}
\label{geo}
\end{figure}

\begin{figure}
\begin{picture}(-300,380)(100,250) 
\put(50,190) { \includegraphics{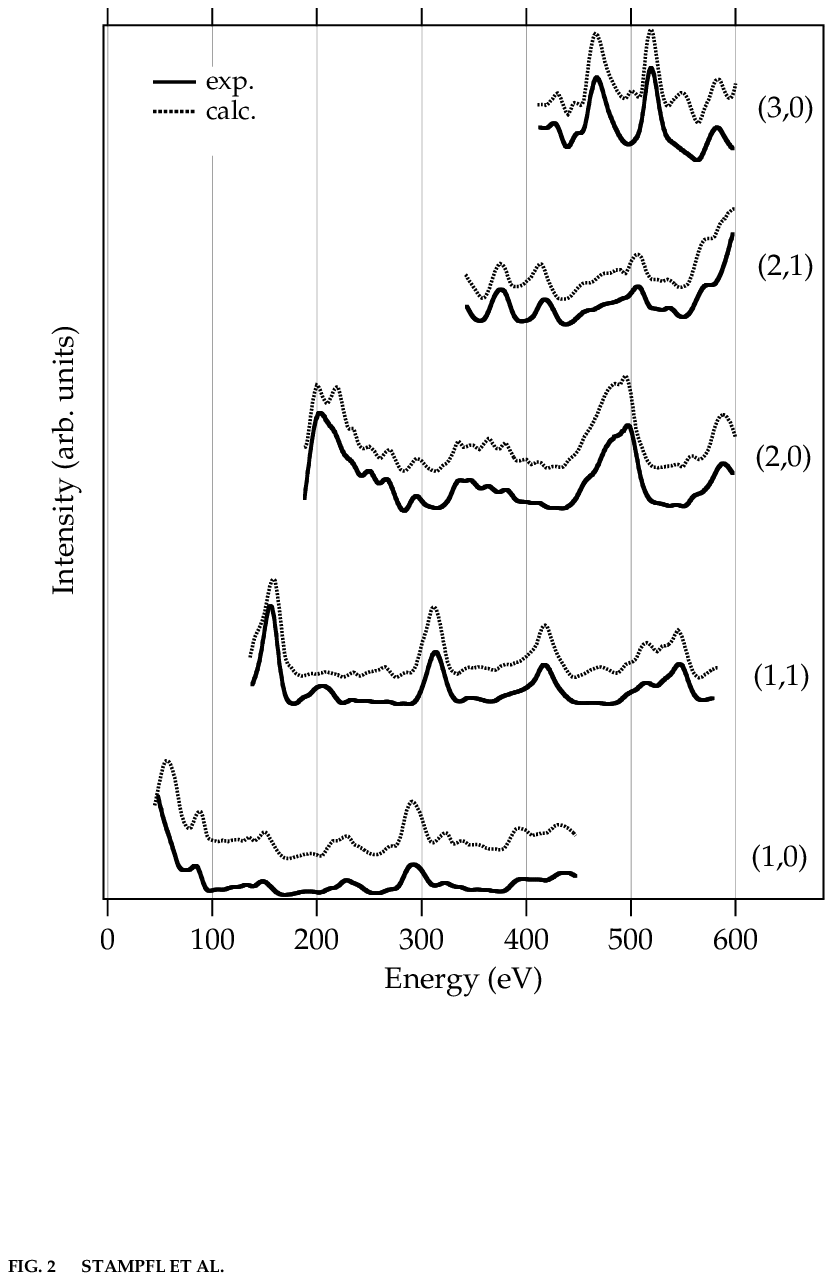}} 
\end{picture}
\caption{Comparison between experimental and calculated LEED 
curves for the best-fit geometry of the (1$\times$1)-O phase on Ru\,(0001). 
The overall $R$-factors are $R_{\rm DE} = 0.21$ and $R_{\rm P} = 0.23$.} 
\label{leed}
\end{figure}

The optimum atomic geometry of the Ru\,(0001)-(1$\times$1)-O phase, as 
provided by the LEED analysis, is displayed in Fig.~\ref{geo}. The intensity
spectra 
for this best-fit geometry are shown in Fig.~\ref{leed}; the corresponding
$R$-factors
are 
$R_{\rm DE} = 0.21$ and $R_{\rm P} = 0.23$.
The Ru-O bond length of 2.00~\AA\,$\pm$0.03~\AA\, is similar to
the data reported previously for the (2$\times$2)-O 
($\Theta = 0.25$) and the (2$\times$1)-O ($\Theta = 0.5$) 
structures which are 2.03~\AA\,  and 2.02~\AA\, respectively 
\cite{pfnur}.
 A remarkable 
feature of the (1$\times$1)-O
 phase is the substantial expansion of the topmost Ru 
layer spacing (2.21~\AA\,)
by 3.7~\%. This expansion is comparable to that derived 
for the O/Zr\,(0001)
system (3.1~\%) where 1\,ML of oxygen is reported to be evenly 
distributed in octahedral sites between the first and second, and second and 
third Zr layers \cite{wang}.

\begin{figure}
\begin{picture}(-300,200)(100,250) 
\put(0,-150) { \includegraphics{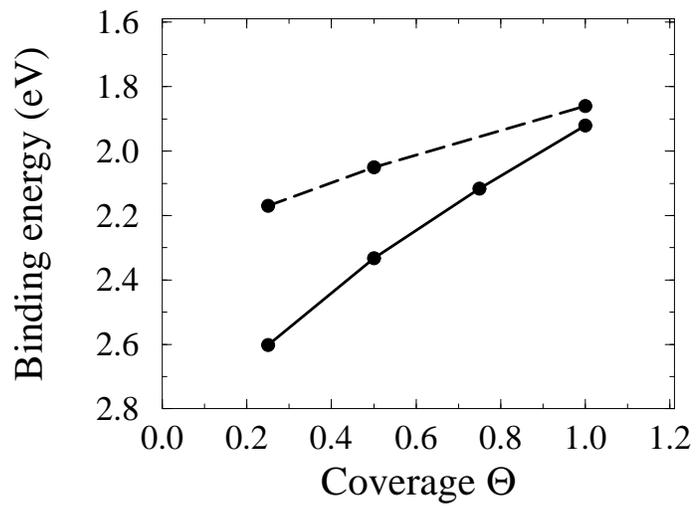}} 
\end{picture}
\caption{Binding energy of O on Ru(0001) at various coverages, with 
respect to   1/2 O$_2$ molecule. 
Adlayers are assumed with fcc-hollow adsorption sites (dashed line)
and with hcp-hollow sites (continuous line).}
\label{energy}
\end{figure}

The experimentally derived structural data agree nicely
with the results of total energy calculations performed for varying
O-coverages, namely  the (2$\times$2)-O, (2$\times$1)-O,
and (1$\times$1)-O structures. For all 
these phases, the energetically most favorable
adsorption site was found to be the 
hcp-site. 
In Tab.~II,
the structural 
parameters for the (1$\times$1)-O system provided by DFT-GGA and LEED are 
compared; the agreement convincingly demonstrates the power 
and reliability of the DFT-GGA calculations.
It is important to notice that the large expansion of the top Ru 
interlayer spacing is not due to subsurface
oxygen. It  results from
the binding of the surface Ru atoms with on-surface oxygen which empties
bonding Ru-Ru $d$ states, weakening the attraction between the top and second
substrate layers.
In Fig.~\ref{energy}, the calculated binding energies for
an oxygen atom in  these three oxygen phases are compiled for the cases where
the O atoms reside in 
hcp and in fcc positions.
It can be seen that with increasing O-coverage the 
binding energy becomes smaller which reflects a repulsive
interaction between the adsorbates and is stronger for oxygen in 
the hcp- than in the fcc-site.
Using these results for a simulation
of thermal desorption spectra \cite{kreuzer} we obtain good agreement with
experimental data \cite{madey} corresponding to the coverage range $\Theta \le
0.5$.
For oxygen in the hcp-site at $\Theta=1$, the energy required
to remove one oxygen atom, i.e. the energy to create an oxygen vacancy in
the adlayer is quite small, namely, 1.2~eV.
This result was estimated from
calculations with a $(2 \times 2)$ surface cell (the artificial
vacancy-vacancy interaction is negligible for our concerns, i.e., less
than $\pm 0.15$ eV, as tested by larger surface cells).
Comparing the binding energies per atom of the two
$\Theta =1$ layers, we find that the energy difference is only
0.06 eV. Because the $\Theta =1$
layer forms subsequently after the completion of the 
lower coverage phases for which the hcp site is by several tenths
of an eV more favorable, it has been  argued~\cite{stampfl}
that, despite the small energy difference, the $\Theta=1$ phase
might be a rather perfect hcp-site adlayer.
Nevertheless, it is obvious that an independent, experimental
structure analysis of this system was mandatory. The very good agreement
between the structural parameters obtained by DFT-GGA and LEED, discussed
above,
provides the required confirmation.

The above results imply that the saturation O-coverage $\Theta \sim 0.5$,
which is reached 
after exposing a Ru\,(0001) surface to O$_{2}$
at low pressure, is only apparent and 
in fact limited by the kinetics of dissociative adsorption. Since the rate of 
impingement increases in proportion to the partial pressure, catalysis under 
high pressure conditions with an excess of O$_{2}$ in the feed becomes rather 
likely to involve surface phases with O-coverages approaching $\Theta = 1$.
Given the fact that the metal-oxygen bond strength is rather weak at the full
coverage (see Fig.~\ref{energy}), it 
is expected to improve the reactivity for reactions of 
the type ${\rm O}_{\rm ad} + {\rm X} \longrightarrow {\rm OX}$.
The present system hence represents, in our opinion, an example for which
extrapolation of the kinetics of a catalytic reaction across the ``pressure
gap''
may be questionable, as also suggested by the quoted work on CO oxidation 
\cite{peden}.

Extrapolation of the data of Fig.~\ref{energy} to even higher coverages
suggests that adsorption of oxygen even beyond $\Theta = 1$ might still be
exothermic. DFT-GGA calculations  at coverages $\Theta = 1.25$
indicate, however,
that further incorporation of O chemisorbed 
on the surface into the (1$\times$1)-O 
structure is unstable.
On the other hand, the occupation of the subsurface octahedral 
adsorption site, just below the first Ru layer, is still exothermic;
it is however, appreciably less favorable than on-surface sites for coverages
$\Theta \leq 1$.
Interestingly, for even deeper positions, the binding energy
of these additional O atoms is similar.
Furthermore, the diffusion energy barriers for oxygen 
penetrating through the first and second Ru layers are found to be such that
the barrier for passing through the second layer is significantly smaller
than that of the first, O-covered Ru layer. 
As a consequence, the theoretical results
suggest that  subsurface oxygen will dissolve into the bulk on entropy grounds. 
>From these calculations (and also from our TDS 
experiments) the existence of a close-packed O overlayer with $\Theta = 3$ on 
Ru\,(0001) as proposed by Malik and Hrbek \cite{malik} can clearly be ruled
out. 
Hence, an impinging NO$_{2}$ molecule gets either reflected by the
(1$\times$1)-O 
surface or dissociation takes place and the oxygen atom is incorporated into 
the subsurface region from where it proceeds into the bulk.

In summary, from a combined investigation using DFT-GGA and LEED we 
confirmed the existence of a high-coverage ($\Theta =1$) phase of
oxygen on Ru\,(0001) and provide a detailed
identification of the geometry.
Furthermore we provided evidence that atomic oxygen may 
enter Ru-bulk once all the on-surface hcp sites are occupied, but not staying 
between the first and second substrate layer. This unusually 
high coverage oxygen structure on Ru\,(0001) is likely to enhance the rate of
oxidation reactions (e.g. of carbon monoxide) and other 
surface reactions involving adsorbed oxygen. It also immediately raises the 
questions of the possible existence of other high coverage surface structures 
likewise achievable by bypassing the ``pressure gap''.

\vspace{1cm}
{\bf Acknowledgment}

The authors are grateful to W.H. Weinberg for stimulating discussions and 
communication of unpublished data and H. Bludau for valuable comments.

\begin{table}
\begin{tabular}{c||c|c}
O-adsorption site & $R_{\rm P}$ & $R_{\rm DE}$ \\
\hline
on top  &  0.68 & 0.47 \\
bridge  &  0.60 & 0.39 \\
fcc     &  0.59 & 0.49 \\
hcp     &  0.23 & 0.21 \\
no oxygen & 0.55 & 0.37 \\
\end{tabular}
\caption{Optimum $R$-factors for different structural models of 
Ru\,(0001)-(1$\times$1)-O.}
\end{table}

\begin{table}
\begin{tabular}{c||c|c}
parameter                     &  DFT-GGA  & LEED   \\
\hline
Ru-O layer spacing (\AA)            & 1.26  & 1.25  $\pm$ 0.02  \\
Ru-O bond length (\AA)              & 2.03  & 2.00  $\pm$ 0.03  \\
first Ru interlayer spacing (\%)       & +\,2.7   &+\,3.7   $\pm$ 1.4   \\
second Ru interlayer spacing (\%)       &$-$\,0.9   & $-$\,0.5   $\pm$ 1.8   \\
\end{tabular}
\caption{Structural parameters of the (1$\times$1)-O phase
 on Ru\,(0001) as obtained by DFT-GGA calculations (see text and ref. 4)
and the LEED analysis.}
\end{table}

\clearpage

\end{document}